# Analysis of Lithium-ion Battery Cells Degradation Based on Different Manufacturers

Ahmed Gailani, Rehab Mokidm, Mo'ath El-Dalahmeh, Ma'd El-Dalahmeh, Maher Al-Greer

# Analysis of Lithium-ion Battery Cells Degradation Based on Different Manufacturers

Ahmed Gailani
School of Computing, Engineering and
Digital Technologies
Teesside University
Middlesbrough, UK
A.Gailani@tees.ac.uk

Ma'd El-Dalahmeh
School of Computing, Engineering and
Digital Technologies
Teesside University
Middlesbrough, UK
Ma'd.El-Dalahmeh@tees.ac.uk

Rehab Mokidm
[1]Battery Energy Storage Devision
Renewable Energy Test Center
Fremont, CA, USA
[2]Fraunhofer ISE
Freiburg im Breisgau, Germany
rehab@retc-ca.com

Maher Al-Greer
School of Computing, Engineering and
Digital Technologies
Teesside University
Middlesbrough, UK
M.Al-Greer@tees.ac.uk

Mo'ath El-Dalahmeh
School of Computing, Engineering and
Digital Technologies
Teesside University
Middlesbrough, UK
Mo'ath.El-Dalahmeh@tees.ac.uk

*Abstract*— **Lithium-ion batteries are recognised as a key technology to power electric vehicles and integrate grid-connected renewable energy resources. The economic viability of these applications is affected by the battery degradation during its lifetime. This study presents an extensive experimental degradation data for lithium-ion battery cells from three different manufactures (Sony, BYD and Samsung). The Sony and BYD cells are of LFP chemistry while the Samsung cell is of NMC. The capacity fade and resistance increase of the battery cells are quantified due to calendar and cycle aging. The charge level and the temperature are considered as the main parameters to affect calendar aging while the depth of discharge, current rate and temperature for cycle aging. It is found that the Sony and BYD cells with LFP chemistry has calendar capacity loss of nearly 5% and 8% after 30 months respectively. Moreover, the Samsung NMC cell reached 80% state of health after 3000 cycles at 35°C and 75% discharge depth suggesting a better cycle life compared to the other two battery cells with the same conditions.**

## I. Introduction

Lithium-ion batteries (LiBs) are seen as the prevalent energy storage technology to power electric vehicles and integrate grid-connected renewable energy sources [1,2]. This is due to their high power and energy density along with a sharp decline in their cost in the recent years [3]. Battery degradation analysis is of critical importance to the profitability of many industrial applications [4]. Moreover, such analysis can determine the applicability of the batteries for second-life usage and recycling [5].

LiB degradation can be quantified when the battery's internal resistance rises and/or capacity fades due to several degradation mechanisms. Some of these degradation mechanisms are the growth of solid electrolyte interphase which causes loss of cyclable lithium [6], the loss of active material caused by mechanical stress and structural changes of electrodes [7], impedance increase [8], and lithium plating [9]. During the cycle life of the battery, the aforementioned degradation mechanisms can be influenced by the depth of discharge (DoD), current rate (C rate) and temperature. During the calendar life, they are affected by the storing temperature and state of charge (SoC).

Previous works studied the effects of different battery parameters and operating conditions on the cycle and calendar life of the battery. In [10], accelerated cycle and calendar aging tests in a period 500 days were conducted on 60 LiB cells to test the impact of different aging parameters on capacity fade and resistance increase. The impact of average state of charge (SoC) and temperature were quantified in the calendar aging test. In the cycle aging test, the DoD and the ampere-hour throughput were quantified while the temperature is constant at 35°C. Then, a mathematical model is created based on the experimental data fitting. In [11], the calendar and cycle aging tests were done for a period of 300 days with the cycle aging impact were quantified for different temperatures and DoDs. In [12], the cycle aging testing conditions includes the effect of C rate on battery state of health (SoH). This is done for a period of 25 days. Similar study is done in [13] to analyse the effects of C rate on LiB cycle aging. This was done for a maximum of 1100 cycles. The SoH estimation techniques for cycle aging in the accelerated aging tests was developed in [14] based on fractional impedance and interval capacity. The maximum equivalent full cycles for the battery was 600 cycles. Other study in [15] quantified calendar aging based on SoC and temperature changes for a period of 470 days while ignoring cycle aging.

The accuracy of any LiB degradation model partly depends on the amount and accuracy of the experimental degradation data. However, the aforementioned studies were limited in terms of either the short time of experimental study used, the accelerated aging conditions that can stimulate faster degradation, or not considering the current developments in LiB materials. Therefore, with these limitations, the generated aging model may be compromised due to data limitation. For instance, in [16], the non-linear model for DoD can be improved if it is based on additional degradation experimental data. As a result, the lifetime projection of the LiB may be compromised. For instance, the current cycle life of the state-of-the-art LiB is expected to last at least 5000 cycles at 40°C temperature and with 20 years calendar time [17].

This paper seeks to improve on earlier research by analysing long-term (2.5 years) cycle and calendar aging data for three different LiB cells based on different manufacturers (BYD, Samsung, and Sony). The collected experimental results provide indications of the lifetime projections of the recent LiB cells and inform modellers and control design. The remainder of the paper is organised as follows: Section II presents the experimental setup of the batteries. Section III presents the results. Section IV offers some concluding remarks.

## II. METHODS

### A. Experimental setup and aging test matrix

Due to LiB cells degrade differently for different chemistries and manufacturers, this study uses cells from different commercial manufactures (BYD, Samsung and Sony) with LFB and NMC chemistries. The cells were cycled using Digatron battery cycler at Fraunhofer Institute for Solar Energy Systems. Extra open circuit voltage and validation tests were done at Renewable Energy Test Center (RETC) in California. The Digatron IBT has six channels with (-100 to +100 A) current range and (0-150 V) voltage range. All the battery cells were connected to environmental chambers of Thermotron type to regulate the temperature during battery cycling. Also, they have been used to control the temperature during calendar aging testing. The experimental setup is shown in Fig.1.

Table. I and Table. II present the cycling and calendar aging tests matrix for the different LiB cells (Sony, BYD, and Samsung (Sam)). For cycling aging, the temperature and the DoD are varied along with the C-rate. For calendar aging, the temperature and the SoC are varied. Three samples of the tested cells are shown in Figs.1-3 for Sony, BYD and Samsung respectively.

TABLE I. CYCLE AGING TEST MATRIX

| Cells type | Capacity (Ah) | Cycling Temperature (°C) | Discharge depth (%) | Ch/Dch current rate |
|---|---|---|---|---|
| Sony 1 | 3 | 25 | (100-60) | 1/1 |
| Sony 2 | 3 | 25 | (60-20) | 1/1 |
| Sony 3 | 3 | 35 | (100-0) | 1/1 |
| Sony 4 | 3 | 35 | (100-0) | 1/1 |
| Sony 5 | 3 | 35 | (80-5) | 1/1 |
| Sony 6 | 3 | 35 | (80-5) | 1/1 |
| BYD1,2 | 25 | 35 | (80-5) | 0.75/0.75 |
| Sam1 | 94 | 20 | (80-20) | 1/1 |
| Sam2 | 94 | 20 | (90-20) | 2/2 |
| Sam3,4 | 94 | 35 | (80-5) | 0.75/0.75 |

TABLE II. CALENDAR AGING TEST MATRIX

| Cells type | Capacity (Ah) | SoC (%) | Temperature (°C) |
|---|---|---|---|
| Sony 7,8 | 3 | 70 | 10 |
| Sony 9,10 | 3 | 70 | 20 |
| Sony 11,12 | 3 | 50 | 35 |
| Sony 13,14 | 3 | 20 | 35 |
| Sony 14,15 | 3 | 70 | 35 |
| BYD 3,4 | 25 | 70 | 10 |
| BYD 5,6 | 25 | 70 | 20 |
| BYD 7,8 | 25 | 20 | 35 |
| BYD 7,8 | 25 | 50 | 35 |
| BYD 9,10 | 25 | 70 | 35 |
| BYD 11,12 | 25 | 100 | 35 |

### B. Cells specifications

Sony US26650FT of type 26650 (cylindrical cell) with dimensions 26.45mm diameter and 65.6mm length is used. The cell chemistry is LiFePO4, with a 3.0 Ah nominal capacity, 3.2 V nominal voltage and 2.0 and 3.6 V cutoff voltages. Fig.2 shows one of the Sony cylindrical tested cell.

BYD cells are of prismatic type with dimensions 173mm*21mm*119.5mm. The cell chemistry is LiFePO4 with a 25 Ah capacity, 3.2 V nominal voltage and 2.0 - 3.8 V lower and upper cutoff voltage respectively. Fig.3 shows a sample of the BYD tested cell.

Samsung SDI AIO 5.5 of prismatic type is used with dimensions 173.2 mm * 45.2 mm * 175.8 mm. The cell chemistry is LiNiMnCo with a 94Ah capacity, and 2.7 V and 4.15 V are the lower and upper cutoff voltage respectively. Fig.4 shows example of the tested cells.

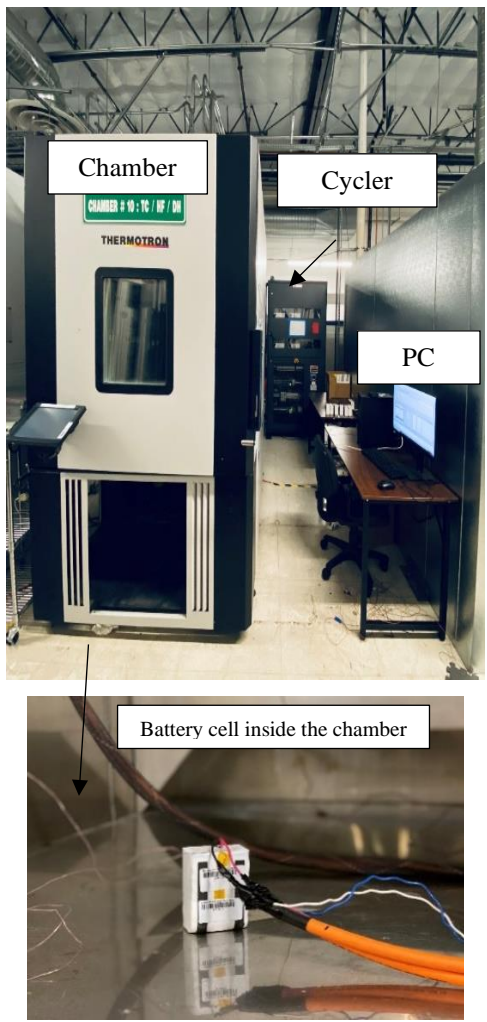

Fig. 1. Battery experimental setup

*C. Cells testing protocols*

At the start of the tests, the cells are left inside the climate chamber at 20°C for 1 hour to ensure thermal equilibrium. Then, the cells are charged with CCCV at 1C rate until the upper cutoff voltage is reached. A rest period of 30 minutes is allowed between charge and discharge. Finally, the cells are discharged at 1C rate until the lower cutoff voltage is reached and hence the initial cell capacity for all the cells are calculated by the cycler by integrating the current.

To measure the capacity after cycling or calendar conditions, a standard reference performance test (RPT) was used with a current of C/20. The RPT was performed each month during the entire testing period. For resistance measurement, a pulse discharge method is used.

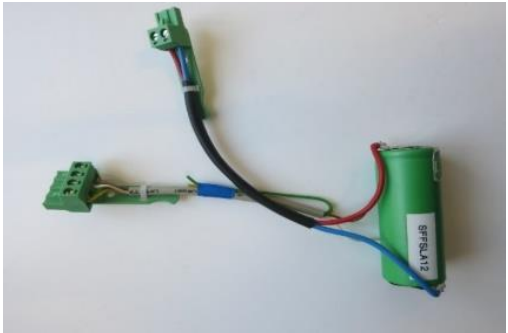

Fig. 2. Sony cylindrical cell

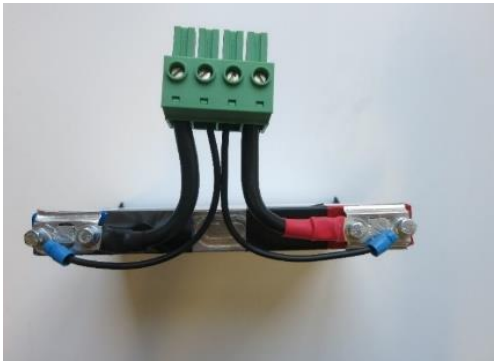

Fig. 3. BYD prismatic cell

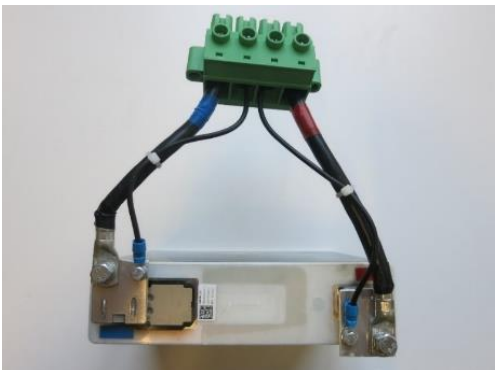

Fig. 4. Samsung Prismatic cell

## III. RESULTS

*A. Calendar aging*

The calendar aging test was done for the Sony and BYD cells since they have the same chemistry (LFP). As seen in Fig.5 and Fig.6, the SoC is kept constant at 70% while the temperature is changed (10°C, 20°C, 35°C) for both cells. The higher the temperature, the higher is the capacity loss due to calendar aging in agreement with the previous literature [4]. After 30 months of calendar aging, the Sony cell capacity loss is 0.33% at 10°C, 1.33% at 20°C and 5.33% at 35°C as in Fig.5. For BYD cell in Fig.6, the capacity loss is 6% at 10°C, 6.4% at 20°C and 10% at 35°C.

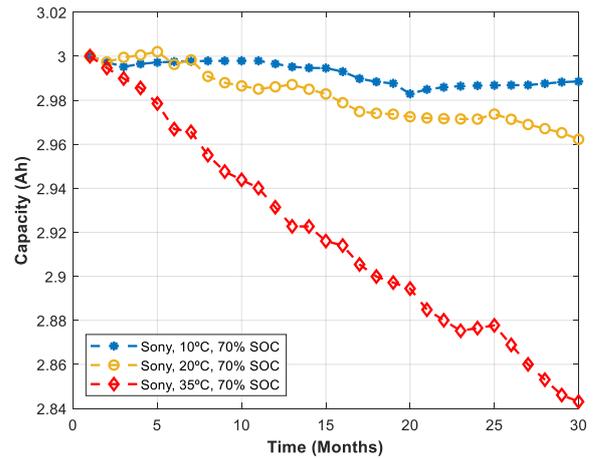

Fig. 5. Capacity fade for Sony cells at different temperatures at 70%SoC

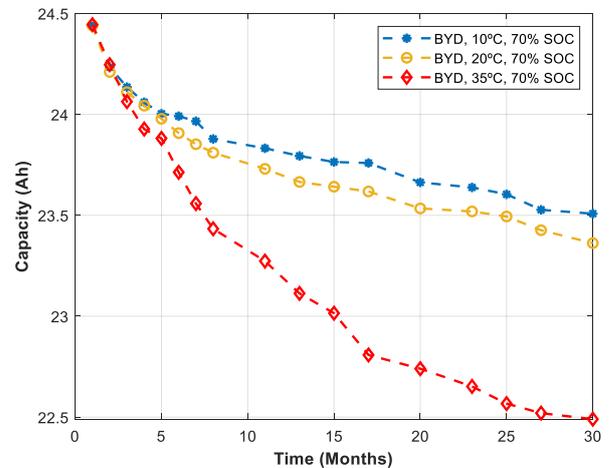

Fig. 6. Capacity fade BYD cells at different temperatures at 70%SoC

The effect of SoC changes while fixed temperature at 35°C is depicted in Fig.7 and Fig.8 for Sony and BYD cells respectively. It can be seen in both Fig.7 and Fig.8 that the lower the SoC level, the lower is the cell capacity losses. For instance, the capacity loss for the Sony cell at 100% SoC is 8% after 30 months compared to just 2.66% at 20% SoC. Similarly, for the BYD cell, the capacity loss at 100% SoC is 12% after 30 months compared to 6.4% at 20% SoC.

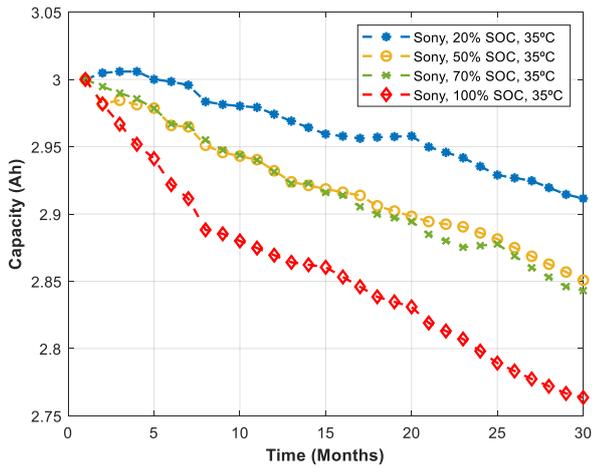

Fig. 7. Capacity fade for Sony cells at different SoC levels at 35°C

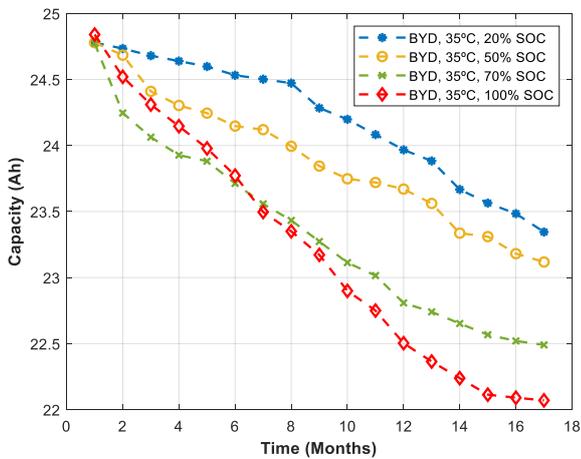

Fig. 8. Capacity fade for BYD cells at different SoC levels at 35°C

The resistance increase for both Sony and BYD cells are shown in Fig.9 and Fig.10 at different temperatures and SoCs respectively. The value of the resistance in the y-axis is the normalised resistance which is the present value divided by the firstly measured resistance value. As shown in Fig.9 and Fig.10, the tested cells exhibit non-linear resistance changes but overall the BYD cells show lower resistance increase as the time passes with 5% increase after 18 months. The maximum resistance increase is 25% for the Sony cell at 35°C after 18 months.

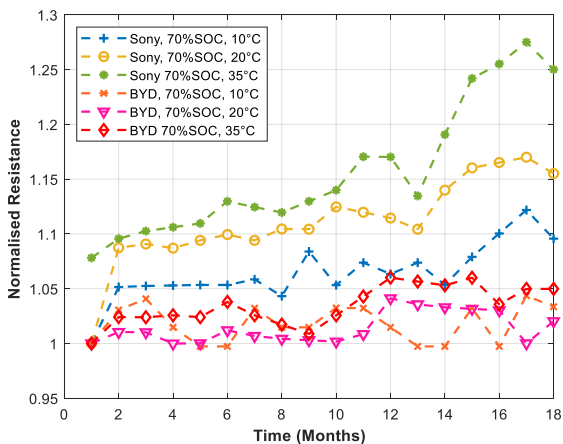

Fig. 9. Resistance increase results for BYD and Sony cells at different temperatures

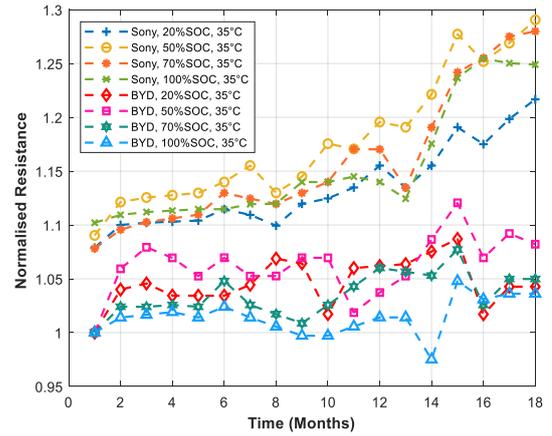

Fig. 10. Resistance increase results for Sony and BYD cells at different charge levels

*B. Cycle aging*

In Fig.11, the Sony, BYD, and Samsung cells were tested at the same condition (75% DoD and T = 35°C). Fig.11 indicates that the Samsung NMC cell has better performance at these conditions with 3000 cycles until it reaches 80% state of health (SoH).

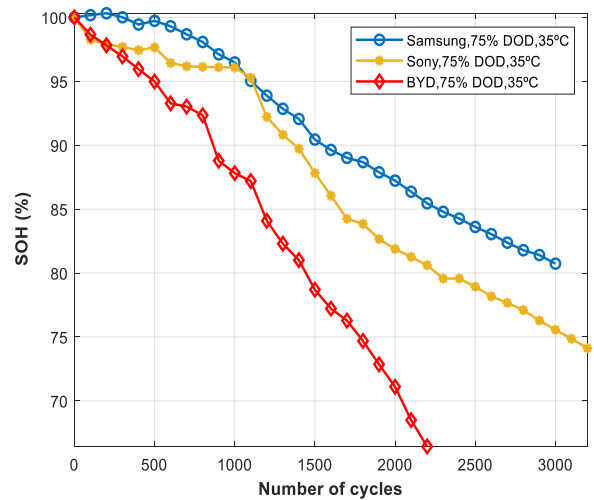

Fig. 11. Capacity fade for different cells at 35°C and 75%DoD

In Fig.12, the impact of different DoDs at 25°C for the Sony cells is quantified. It can be seen that discharging from 100% SoC to 60% SoC has lower capacity losses than discharging from 60% to 20% SoC. Moreover, the impact of rising the temperature by 10°C for the same Sony cell is greater on the capacity loss than the level of DoD. The capacity fade for the Samsung cells at 1 to 2 C rates at 20°C is shown in Fig.13 where the higher the C-rate the higher is the capacity losses. At 1C rate the Samsung cell is expected to have more than 3000 cycle before reaching 70% SoH.

In Fig.14, two Samsung cells with the exact same conditions at 75% DoD and 0.75 C rate were tested to investigate cell to cell variations. It can be seen that the two cells behavior is nearly identical with 78% SoH after 3000 cycles.

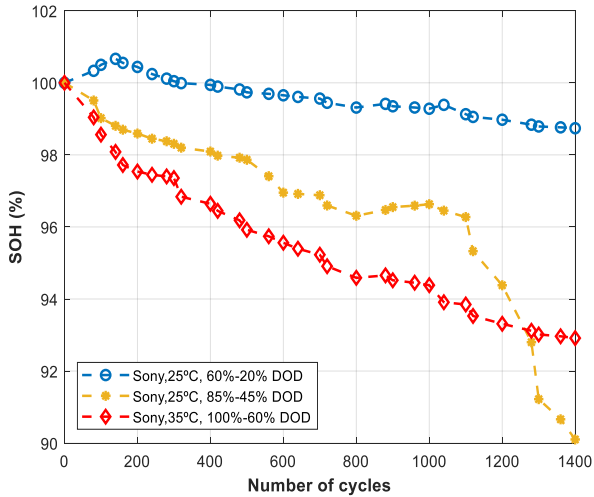

Fig. 12. Capacity fade for Sony cells 40%DoD and temperatures 25-35°C

In Fig.15, the resistance increase for all the three cells was measured at the same temperature (35°C) and different DoDs. In general, the Samsung cell exhibited lower resistance increase compared to the other two. Moreover, DoD level has less effect on battery resistance increase compared to capacity fade at the same conditions. For instance, the Samsung cell's resistance increase is nearly 10% after 3000 cycles compared to 25% for the Sony cell. The BYD cell's resistance increase is 78% after 2000 cycles.

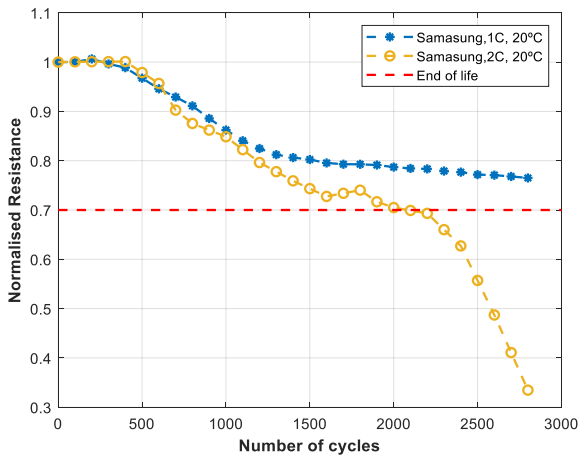

Fig. 13. Capacity fade for Samsung cell at different current rates at 20°C

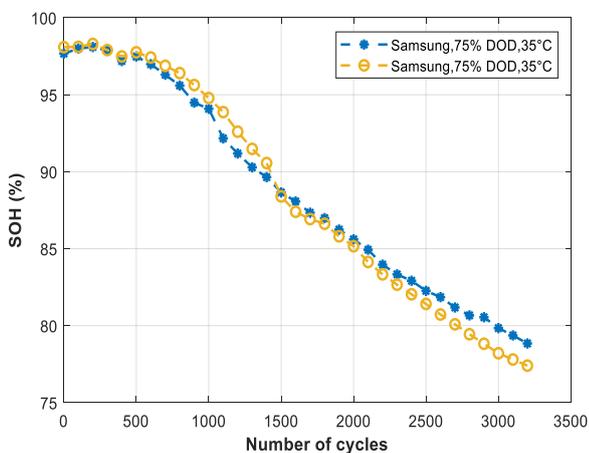

Fig. 14. Capacity fade for Samsung cells at different current rates at 30°C

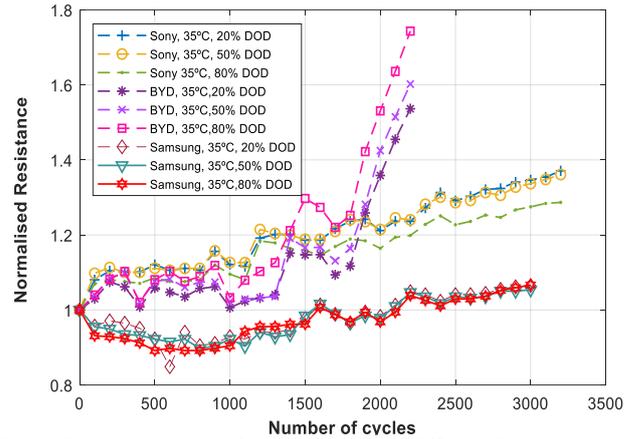

Fig. 15. Resistance increase for different cells at different discharge depths

The battery degradation results have three potential implications. First, the results serve as a benchmark for battery techno-economic studies by improving battery degradation [18]. Second, the results present an opportunity for an accurate LiB degradation modelling based on long-term data and different chemistries (LFP and NMC). Third, the results show that degradation trends are different between battery cells from the same or different manufacturer which necessitates detailing all the battery cell information in future research for accurate degradation assessment.

IV. CONCLUSION

This paper presented a calendar and cycle aging analysis in terms of capacity fade and resistance increase for LiB cells from different manufacturers. The cells are 3Ah LFP from Sony, 25Ah LFP from BYD and 94Ah NMC from Samsung. The cells were tested for calendar aging where SoC and temperature were varied and cycle aging where C rate, DoD and temperature were varied. It is found that the temperature is the main parameter that affect capacity fade and resistance increase for both calendar and cycle aging. After 30 months of calendar aging testing, the maximum cell capacity loss is 5.33% and 8% at 35°C for both Sony and BYD cells respectively. However, after 18 months, the resistance increase at the same conditions was found to be 5% and 25% for the BYD and Sony cells respectively. For cycling aging, it is found that the Samsung NMC cell outperforms the Sony and the BYD cells at the same 1C rate, 35°C temperature and 70% DoD conditions reaching 80% SoH after 3000 cycles. It was also found that the expected life cycle for the Samsung cells is nearly 3000 cycle at 1C and around 2000 cycles at 2 C.

ACKNOWLEDGEMENT

Experimental data presented in this work were collected by the second author (Rehab Mokidm) at Fraunhofer Institute for Solar Energy Systems and Renewable Energy Test Center at California.

REFERENCES

[1] A. Yang, Y. Wang, K. L. Tsui, and Y. Zi, "Lithium-ion Battery SOH Estimation and Fault Diagnosis with Missing Data," in 2019 IEEE International Instrumentation and Measurement Technology Conference (I2MTC), 2019, pp. 1-6.

[2] B. Xu, A. Oudalov, A. Ulbig, G. Andersson, and D. S. Kirschen, "Modeling of Lithium-Ion Battery Degradation for Cell Life

Assessment," IEEE Transactions on Smart Grid, vol. 9, no. 2, pp. 1131-1140, 2018.

[3] BloombergNEF, "New Energy Outlook," 2018. Accessed: 3 April 2019. [Online]. Available: https://bnef.turtl.co/story/neo2018?teaser=true.

[4] A. Gailani, M. Al-Greer, M. Short, and T. Crosbie, "Degradation Cost Analysis of Li-Ion Batteries in the Capacity Market with Different Degradation Models," Electronics, vol. 9, no. 1, 2020.

[5] G. Harper et al., "Recycling lithium-ion batteries from electric vehicles," Nature, vol. 575, no. 7781, pp. 75-86, 2019.

[6] X.-G. Yang, Y. Leng, G. Zhang, S. Ge, and C.-Y. Wang, "Modeling of lithium plating induced aging of lithium-ion batteries: Transition from linear to nonlinear aging," Journal of Power Sources, vol. 360, pp. 28-40, 2017.

[7] M. S. D. Darma et al., "The influence of cycling temperature and cycling rate on the phase specific degradation of a positive electrode in lithium ion batteries: A post mortem analysis," Journal of Power Sources, vol. 327, pp. 714-725, 2016.

[8] S. E. Li, B. Wang, H. Peng, and X. Hu, "An electrochemistry-based impedance model for lithium-ion batteries," Journal of Power Sources, vol. 258, pp. 9-18, 2014.

[9] U. R. Koleti, T. Q. Dinh, and J. Marco, "A new on-line method for lithium plating detection in lithium-ion batteries," Journal of Power Sources, vol. 451, p. 227798, 2020.

[10] J. Schmalstieg, S. Käbitz, M. Ecker, and D. U. Sauer, "A holistic aging model for Li(NiMnCo)O2 based 18650 lithium-ion batteries," Journal of Power Sources, vol. 257, pp. 325-334, 2014.

[11] Smith, A. Saxon, M. Keyser, B. Lundstrom, Z. Cao, and A. Roc, Life prediction model for grid-connected Li-ion battery energy storage system. 2017, pp. 4062-4068.

[12] S. Barcellona and L. Piegari, "Effect of current on cycle aging of lithium ion batteries," Journal of Energy Storage, vol. 29, p. 101310, 2020.

[13] A. Yang et al., "A comprehensive investigation of lithium-ion battery degradation performance at different discharge rates," Journal of Power Sources, vol. 443, p. 227108, 2019.

[14] Q. Yang, J. Xu, X. Li, D. Xu, and B. Cao, "State-of-health estimation of lithium-ion battery based on fractional impedance model and interval capacity," International Journal of Electrical Power & Energy Systems, vol. 119, p. 105883, 2020.

[15] J. Schmitt, A. Maheshwari, M. Heck, S. Lux, and M. Vetter, "Impedance change and capacity fade of lithium nickel manganese cobalt oxide-based batteries during calendar aging," Journal of Power Sources, vol. 353, pp. 183-194, 2017.

[16] M. S. Javadi, A. Anvari-Moghaddam, and J. M. Guerrero, "Optimal scheduling of a multi-carrier energy hub supplemented by battery energy storage systems," in 2017 IEEE International Conference on Environment and Electrical Engineering and 2017 IEEE Industrial and Commercial Power Systems Europe (EEEIC / I&CPS Europe), 2017, pp. 1-6.

[17] J. E. Harlow et al., "A Wide Range of Testing Results on an Excellent Lithium-Ion Cell Chemistry to be used as Benchmarks for New Battery Technologies," Journal of The Electrochemical Society, vol. 166, no. 13, pp. A3031-A3044, January 1, 2019.

[18] A. Gailani, M. Al-Greer, M. Short, T. Crosbie, and N. Dawood, "Lifetime Degradation Cost Analysis for Li-Ion Batteries in Capacity Markets using Accurate Physics-Based Models," Energies, vol. 13, no. 11, 2020.